\documentstyle[emulateapj]{article}
\def\Meszaros{M\'esz\'aros~}

\begin{document}

\title{GeV photons from up-scattering of supernova shock breakout X-rays by an outside GRB jet }

\author{Xiang-Yu Wang\altaffilmark{1,2} and Peter M\'esz\'aros\altaffilmark{1,3}}
\altaffiltext{1}{Department of Astronomy and Astrophysics,
Pennsylvania State University, University Park, PA 16802, USA;
xywang@astro.psu.edu, nnp@astro.psu.edu}

\altaffiltext{2}{Department of Astronomy, Nanjing University,
Nanjing 210093, China} \altaffiltext{3}{Center for Gravitational
Wave Physics and Department of Physics, Pennsylvania State
University, University Park, PA 16802, USA}

\begin{abstract}

Shock breakout X-ray emission has been reported for the first time
from a supernova connected with a gamma-ray burst, namely
GRB060218/SN2006aj. The gamma-ray emission and the power-law
decaying X-ray afterglow  are ascribed to a highly relativistic
jet, while the thermal soft X-rays are thought to be produced when
the radiation-dominated shock breaks from the optically thick
stellar wind. We study the inverse Compton emission of the
breakout thermal soft X-rays scattered by relativistic electrons
in the  jet forward shock,  which  is expected to be at larger
radii than the breakout shock.  This IC emission produces sub-GeV
to GeV photons, which may be detectable by GLAST. The detection of
such GeV photons simultaneously with the supernova shock breakout
emission would give evidence for the presence of a GRB jet ahead
of the shock while the shock is breaking out. The anisotropic
scattering between the X-rays and relativistic electrons may lead
to large angle emission outside of the jet opening angle. This has
implications for the detection of GeV photons from burstless type
Ib/c hypernova shock breakout, which due to its more isotropic
emission might be observed with  wide-field X-ray cameras such as
{ LOBSTER}.

\end{abstract}



\section{Introduction}

Recent observations of GRB060128/SN2006aj (Modjaz et al. 2006;
Sollerman et al. 2006; Pian et al. 2006; Mazzali et al. 2006),
aside from providing new evidence that gamma-ray bursts (GRBs) are
connected with supernovae (SNe), also for the first time provide
direct evidence for  shock breakout soft thermal X-ray emission
(Campana et al. 2006).  The gamma-rays are, as usual, thought to
arise from the internal dissipation of ultra-relativistic jets.
The power-law decaying afterglow at times $>10^4$ sec is
consistent with this standard picture of GRBs (e.g. Zhang \&
\Meszaros 2004; Piran 2005). The thermal soft X-ray emission is
interpreted as arising from the breakout of a radiation-dominated
shock, which could be driven by  the cocoon of the relativistic
jet (M\'esz\'aros \& Rees 2001; Ramirez-Ruiz et al. 2002; Zhang et
al. 2003), the slower outer parts of the jet, or the outermost
parts of the envelope which get accelerated to a mildly
relativistic velocity (Colgate 1974; Tan et al. 2001). The
relativistic jet accelerates after leaving the supernova
photosphere and runs ahead of the shock while the less
relativistic shock is breaking out from the stellar wind. The
observed breakout X-ray emission continues after the end of the
GRB prompt emission and last until the early afterglow phase. It
contains two components, a thermal component and a non-thermal
one. The thermal component becomes increasingly dominant during
the first ~3000 sec after the trigger of this GRB. Since the
relativistic GRB jet is expected to be located outside of the
breakout shock, we point out here that the incoming shock X-rays
will be upscattered by the non-thermal electrons accelerated in
the afterglow forward shock to GeV energies.

Gamma-ray emission from bremsstrahlung and inverse-Compton
emission of thermal electrons during the supernova shock breakout
was suggested by Colgate (1974).  Tan et al. (2001) argued that
gamma-rays of GRB980425 associated with SN1998bw are produced by
the interaction of a mildly relativistic ejecta, which results
from the acceleration of the supernova shock, interacting with a
dense circumstellar wind. Waxman \& Loeb (2001) suggested that
during the shock breakout of type II supernovae, the shock may
transition to a collisionless one and accelerate non-thermal
electrons which might produce hard X-ray flares. In these
scenarios, no GRB jets are required, leaving a power-law decaying
X-ray afterglow still in need of interpretation. The strong soft
X-ray thermal photon bath and the compact size of the progenitor
star of type Ib/c supernova would combine to make a large pair
production absorption optical depth for GeV photons if they could
be produced in this scenario. This is different from the case of
scattering by the jet forward shock electrons, in which GeV
photons do not suffer significant absorption because the jet has a
relativistic motion and reached a much larger radius. We suggest
that detection of GeV emission accompanying supernova shock
breakout will favor the normal scenario that involves a highly
relativistic jet to produce the gamma-ray emission in
supernova-connected GRBs like GRB060218/SN2006aj.

In section 2, we first study the emission signature of this IC
process and then consider the anisotropic scattering effect and
pair production absorption effect on the received IC flux in
section 3 and 4 respectively. Finally we give the conclusions.

\section{Inverse Compton scattering of the SN shock breakout X-rays by afterglow shock electrons}
We consider the inverse Compton (IC) scattering of shock breakout
X-rays after the end of the prompt gamma-ray emission but before
the sharp X-ray decay, namely from ~1000 sec to ~3000 sec after
the GRB trigger. During this time, the X-ray emission detected by
{\it Swift} XRT can be divided into two components, one being a
thermal component and the other a non-thermal one (Campana et al,
2006). The thermal X-ray component becomes increasingly dominant
and has a luminosity of $\sim10^{46} \rm erg s^{-1}$. After $\sim
3000$ s, the X-ray emission undergoes a sharp decay and an X-ray
afterglow decaying as a power law in time emerges later. The
underlying afterglow flux extrapolated from the late power-law
decaying afterglow is more than 30 times lower than the thermal
X-ray emission, and is not enough to account for the non-thermal
emission of the previous phase. This early non-thermal component
could come from the  shock breakout itself or from the extended
internal shock emission.

The magnetic field energy density in the forward shock frame is
lower than that of the breakout X-ray emission if the
equipartition factor $\epsilon_B<0.1L_{X,46}E_{50}^{-1}t_3$
(hereafter $C_n=C/10^n$ in c.g.s. units). This implies that the
shock breakout X-ray emission is the dominant cooling source for
the early afterglow forward shock electrons. Below we will show
that the strong shock breakout X-ray emission will cause the
afterglow electrons to be in the fast cooling regime, and
therefore most energy of the shocked electrons during this period
goes into the IC emission.

The gamma-ray emission of GRB060218 is weak compared to typical
GRB with an isotropic gamma-ray energy $E_\gamma\simeq10^{50}{\rm
erg} $. Here we consider a jet with isotropic kinetic energy
$E=10^{50}{\rm erg}$ expanding in a surrounding wind medium with
density profile
\begin{equation}
\rho=K r^{-2}, K\equiv\frac{\dot{M}}{4\pi v_w},
\end{equation}
where $\dot{M}$ is the mass-loss rate and $v_w$ is the wind
velocity. From energy conservation of the adiabatic shock, we can
get  the Lorentz factor and radius of the afterglow shock
(Chevalier \& Li 1999),
\begin{equation}
\Gamma=4.3 E_{50}^{1/4}\dot{m}^{-1/4}t_3^{-1/4},
\end{equation}
\begin{equation}
R=1.7\times10^{15}E_{50}^{1/2}\dot{m}^{-1/2}t_3^{1/2} \rm cm
\end{equation}
where $\dot{m}\equiv(\dot{M}/10^{-5}M_{\sun} {\rm
yr^{-1}})/(v_w/10^3 \rm km s^{-1})$

From $L_X=4\pi R^2\Gamma^2U'_Xc$, we have that the energy density
of shock breakout X-ray photons in the forward shock frame is
$U'_X=L_X/(4\pi\Gamma^2R^2c)$, where $L_X=10^{46}{\rm erg s^{-1}}$
is the  observed luminosity of the shock breakout X-rays. The
Lorentz factor of the electrons which cool by IC scattering of
shock breakout X-rays  in the dynamical time, $R/\Gamma c$, is
\begin{equation}
\gamma_c=\frac{3\pi m_e c^3 \Gamma^3 R}{\sigma_T L_X}= 6
L_{X,46}^{-1}E_{50}^{7/8}\dot{m}^{-7/8}t_3^{-1/4},
\end{equation}
 while the minimum Lorentz factor of the
post-shock electrons is
\begin{equation}
\gamma_m=\bar\epsilon_{e}
\frac{m_p}{m_e}(\Gamma-1)\simeq610\bar\epsilon_{e,-1}E_{50}^{1/4}{\dot{m}}^{-1/4}t_3^{-1/4}
\end{equation}
where $\bar\epsilon_{e}$ is the usual equipartition factor
($\epsilon_e$) of shocked electrons multiplied by a factor of
$(p-2)/(p-1)$ ($p$ is the energy distribution index of the
electrons). So if the break out emission has a luminosity larger
than
\begin{equation}
L_{X}>2\times10^{44}\bar\epsilon_{e,-1}^{-1}E_{50}^{5/8}{\dot{m}}^{-5/8}
\rm erg  s^{-1},
\end{equation}
one obtains $\gamma_m>\gamma_c$, which implies that most of the
energy of the afterglow shocked electrons during the shock
breakout period would go into the IC emission.

As the forward shock propagates in the surrounding medium, the
energy that goes into the newly shocked electrons per unit
observer's time is  $L_e=2\times10^{46}\epsilon_{e,-1}
E_{50}t_3^{-1} \rm erg$ (Wang, Li \& \Meszaros 2006). The
bolometric luminosity $L_{IC}$ of the IC emission is equal to
$L_e$, so the total energy loss of the shocked electrons during
the time from 1000 sec to 3000 sec is about $ E_{IC}\simeq
2\times10^{49} \epsilon_{e,-1}E_{50} \rm erg$ and the total IC
emission fluence is
\begin{equation}
{\Psi}_{IC}=10^{-5}\epsilon_{e,-1}E_{50}(\frac{D}{145 \rm
Mpc})^{-2} \rm erg cm^{-2},
\end{equation}
where $D=145{\rm Mpc}$ is the distance of GRB060218/SN2006aj. The
observed IC $\nu F_\nu$ flux peaks at
\begin{equation}
\varepsilon_{IC,p}\simeq
2\gamma_m^2(3kT)=0.2\bar\epsilon_{e,-1}^2E_{50}^{1/2}\dot{m}^{-1/2}t_3^{-1/2}(\frac{kT}{0.1\rm
KeV}) \rm GeV
\end{equation}
where $kT\simeq0.16{\rm KeV}$ is the black body temperature of the
thermal spectrum of the shock breakout emission (Campana et al.
2006). The IC energy spectrum($\nu F_{\nu}$) has indices of $1/2$
and $-(p-2)/2$ before and after the break at $\varepsilon_{IC,p}$
respectively. The upcoming GLAST LAT detector has an effective
detection area of $10^4\rm cm^2$, so it can detect the sub-GeV to
GeV photons with the above flux, and may even identify the peak
energy of this IC emission.

The jet afterglow electrons may also produce high energy emission
through the self-synchrotron Compton process (Zhang \& \Meszaros
2001). However, based on the assumption that the extrapolated
synchrotron flux during the early afterglow is $>30$ times lower
than the shock breakout X-ray flux{\footnote{The real contribution
to the X-ray flux by the synchrotron emission would be suppressed
due to the enhanced cooling of electrons by the shock breakout
X-ray emission (Wang et al. 2006).}}, the self-synchrotron Compton
flux is then lower by the same factor.

After the shock breaks out of the optically thick wind, its
emission drops very sharply, and the jet afterglow emission
becomes dominant. The remaining energy, $E-E_{IC}$, of the jet
continues to power the power-law decaying afterglow at later
times. The jet is likely to enter into the sub-relativistic phase
around $t\sim 10^5-10^6$ sec for a low burst energy, and  the
power decay rate should transition from $t^{\frac{2-3p}{4}}$ for
the relativistic phase (Dai \& Lu 1998; Chevalier \& Li 1999) to
$t^{\frac{8-7p}{6}}$ for the sub-relativistic one (Waxman 2004)
when the cooling frequency is below the X-ray band. However, for
$p\sim 2$, the decay index almost does not change, and is
consistent with the observed single decay rate $F_X\propto
t^{-1.15\pm0.15}$ (Cusumano 2006).

\section{Anisotropic inverse Compton scattering effect on the received IC flux and the case of off-axis jets}
The shock breakout emission comes from the region of the immediate
stellar wind surrounding the SN/GRB progenitor at
$R\sim5\times10^{12}\rm cm$ (Campana et al. 2006), which is much
smaller than the afterglow shock radius, so the shock breakout
X-ray photons  move outward almost radially, viewed from the
afterglow region. The seed photons are anisotropic seen by the
isotropically distributed electrons in the forward shock, so the
scatterings between photons and relativistic electrons are
anisotropic and there would be more head-on scatterings (e.g.
Ghisellini 1991). In the comoving frame of the forward shock, the
IC emission power depends on the relative angle $\theta_s$ between
the scattered photon direction and the seed photon beam direction
as
\begin{equation}
P(\theta_s)d\theta_s\propto (1- {\rm
cos}\theta_s)^{\frac{q+1}{2}}{\rm sin}\theta_s d\theta_s
\end{equation}
(e.g. Brunetti 2000;  Fan \& Piran 2006), where $q$ is the
power-law energy distribution index of  the scattering electrons.
For fast-cooling of relativistic electrons in our case,
$q=p+1\simeq 3$. The factor $(1-{\rm
cos}\theta_s)^{\frac{q+1}{2}}$ results in more power emitted at
large angles relative to the seed photon beam direction. By simple
algebraical calculation, one can obtain the fraction of photons
scattered into the angles $0\la\theta_s\la\pi/2$ in the shock
frame (corresponding to the $1/\Gamma$ cone of the relativistic
afterglow jet in the observer frame due to aberration of light),
i.e.
\begin{equation}
f_{s}\simeq\frac{\int_0^{\pi/2}(1-{\rm cos}\theta_s)^2{\rm
sin}\theta_s d\theta_s}{\int_0^{\pi}(1-{\rm cos}\theta_s)^2{\rm
sin}\theta_s d\theta_s}=\frac{1}{8}.
\end{equation}

Similarly, one can get that the fraction of photons falling into
the cone of $2/\Gamma$ (corresponding to $\theta_s=2 {\rm arctg}
2$) is $1/2$.  So the effect of increased head-on scattering can
decrease the IC emission in the $1/\Gamma$ cone along the
direction of the photon beam, but enhance the emission at larger
angles, if the jet has an opening angle $\theta_j\la 1/\Gamma$.
However if the opening angle is  $\theta_j\gg1/\Gamma$, the jet
geometry can be approximately regarded  as a sphere, and the
observed IC power after integration over angles  should be the
same  in every direction  in the observer frame (Wang, Li \&
\Meszaros 2006; Fan \& Piran 2006).  So, the received IC fluence
will no longer be reduced. At $t\sim10^3$ sec, the jet Lorentz
factor is $\Gamma=4.3 E_{50}^{1/4}\dot{m}^{-1/4}$. So for typical
mass-loss rate with $\dot{m}=1$ and a weak burst such as GRB060218
with energy $E=10^{50}\rm erg$,  a jet with an opening angle
$\theta_j>1/\Gamma\simeq0.25$ is needed to avoid significant
reduction of the received IC fluence. Considering that the jet has
a low energy, such an opening angle may not be unreasonable.

An interesting phenomenon arises when the jet opening angle is not
larger than $1/\Gamma$ (i.e. $\theta_j\simeq 1/\Gamma$) and the
observer lies outside of the cone of the GRB jet, e.g. lies
between $1/\Gamma$ and $2/\Gamma$, the probability of which may be
somewhat larger than the case within the $1/\Gamma$ cone. The
observer would miss the GRB, but may still detect the IC emission
associated with the SN shock breakout because the prompt gamma-ray
flux decreases very sharply as
$[\Gamma_0(\theta_{obs}-\theta_0)]^{-6}$ while the anisotropic IC
emission decreases  modestly with the relative angle (almost
constant within $2/\Gamma$), where $\Gamma_0$ is the jet initial
Lorentz factor, $\theta_{obs}$ and $\theta_0$ are the observing
angle relative to the jet axis and jet opening angle respectively.
The break out X-ray emission could be detected by wide-field X-ray
detectors such as the future LOBSTER mission{\footnote{see
http://www.src.le.ac.uk/projects/lobster/}} (Calzavara \& Matzner
2004) even if there is no trigger in a GRB detector. LOBSTER would
have the capability of detecting the shock breakout X-ray flash
from type II supernova explosions up to 100Mpc, as the shocked gas
of type II supernova is predicted to emit $\sim10^{45}{\rm erg
s^{-1}}$ with an effective temperature of $2\times10^5$ K (Klein
and Chevalier 1978, Ensman and Burrows 1992). For shock breakout
from type Ib/c supernovae like SN2006aj, the soft X-ray flux is
larger by about two orders of magnitude, so LOBSTER could detect
this kind of explosions up to $\sim1$ Gpc. The IC emission will be
delayed relative to the shock break out emission due to a longer
light travel distance. It is important to note that this IC GeV
emission will not be polluted by other possible high energy
emission processes from the jet itself.

Another nearby supernova connected GRB, GRB031203/SN2003lw, which
is a  similarly weak burst (Sazonov et al. 2004; Soderberg et al.
2004; Watson et al. 2004), has been suggested to be a possible
off-axis burst (Ramirez-Ruiz et al. 2005), i.e. the line of sight
of the observer is outside of the jet opening angle. If this type
Ic SN also has a shock breakout soft X-ray emission like
GRB060218/SN2006aj, which is possibly not observed due to the
energy threshold of the {\it INTEGRAL} detector, GeV photons
produced by the upscattering of these X-rays by relativistic
electrons in the jet forward shock are expected to have a fluence
detectable with GLAST, considering that the on-axis isotropic
energy of the jet could be $E\sim10^{53}{\rm erg}$ (Ramirez-Ruiz
et al. 2005).

\section{Pair production opacity for high energy photons}

Here we consider the pair production opacity and the high energy
photon cutoff due to the absorption by the  shock breakout thermal
photons. First we calculate the optical depth for high energy
photons that  originate from some possible non-thermal process
occurring in the shock breakout itself. The optical depth for high
energy photons, $\varepsilon_\gamma=2 (m_e
c^2)^2/\varepsilon_X=5(kT/0.1{\rm keV})^{-1}{\rm GeV}$,
annihilating with thermal photons of characteristic energy
$\varepsilon_X=kT$  is
\begin{equation}
\begin{array}{ll}
\tau_{\gamma\gamma}=0.1\sigma_T (\frac{L}{4\pi R^2 c
kT})(\frac{R}{\eta})\\
=10^5 L_{46}(\frac{R}{5\times10^{12}{\rm
cm}})^{-1}(\frac{\eta}{20})^{-1}(\frac{kT}{0.1{\rm keV}})^{-1}
\end{array}
\end{equation}
where ${R}=5\times10^{12}{\rm cm}$ is the inferred radius of the
breakout shock, $R/\eta$ is shocked shell thickness. Considering
the exponential tail of the Plank function of the thermal
spectrum, this very large optical depth will induce a cutoff at
\begin{equation}
E_{cut}\simeq\varepsilon_\gamma/{\rm
ln}(\tau_{\gamma\gamma})\simeq0.4(kT/0.1 \rm keV)^{-1} \rm GeV .
\end{equation}

Then we calculate the optical depth for high energy photons
produced through the up-scattering of breakout X-rays by afterglow
shock electrons. The energy of the photons annihilating with the
thermal photons with characteristic energy $\varepsilon_X=kT$ is
$\varepsilon_\gamma=2 \Gamma^2 (m_e
c^2)^2/\varepsilon_X\simeq90(\Gamma/4.3)^2(kT/0.1{\rm
keV})^{-1}{\rm GeV}$, where $\Gamma$ is the bulk Lorentz factor of
the afterglow shock. The pair production absorption optical depth
for this energy is
\begin{equation}
\tau_{\gamma\gamma}=0.1\sigma_T
(\frac{U'_X}{kT/\Gamma})(\frac{R}{\Gamma})
=400L_{46}E_{50}^{-1}\dot{m} (\frac{kT}{0.1{\rm keV}})^{-1}.
\end{equation}
In this case, for representative parameter values, the cutoff
energy is
\begin{equation}
E_{cut}\simeq\varepsilon_\gamma{\rm
ln}(\tau_{\gamma\gamma})\simeq15
E_{50}^{1/2}\dot{m}^{-1/2}t_3^{-1/2}(kT/0.1 \rm keV)^{-1}\rm GeV .
\end{equation}

Thus, we conclude that, at the radius of the shock breakout, only
a negligible fraction of $\sim \rm GeV$ photons can escape from
the strong thermal photon bath, while at the larger radius of the
afterglow shock, up to $\sim10$ GeV photons can escape the pair
production absorption{\footnote{Very high energy photons, say
$>10$TeV photons, may not have large pair production absorption,
but the flux is too low to be detectable except when the spectrum
is very flat.}}. This implies that future detections of GeV
photons accompanying the SN shock breakout in GRBs like
GRB060218/SN2006aj can provide evidence that there exists a
relativistic jet ahead of the supernova shock break out.

\section{Conclusions}
The large energy and thermal spectrum of the  soft X-ray emission
observed in the early stages of a nearby GRB/SN,
GRB060218/SN2006aj, supports the view that this arises from  a
radiation-dominated shock which is breaking out of the strong
stellar wind. The prompt gamma-ray emission and the power law
decaying afterglow can be interpreted conventionally as a highly
relativistic jet, which presumably has moved to larger radii while
the shock is breaking out. In this paper we have suggested that
the upscattering of the shock breakout soft X-rays by relativistic
electrons in the GRB jet forward shock would produce sub-GeV to
GeV photons that could be detected by GLAST.

It is possible that some nonthermal processes in the breakout
shock itself could also produce GeV emission. However, in this
case, the photons with energies above a few hundreds of MeV cannot
escape the absorption by the abundant soft photons, and will be
undetectable. Thus,  detection of high energy photons with
energies above GeV accompanying the shock breakout of GRBs/SNe
like GRB060218/SN2006aj will favor the scenario involving a
relativistic jet producing the GRB, which outruns the slower
supernova shock.

We also discussed the anisotropic IC scattering effect on the
received GeV fluence. When the jet opening angle $\theta_0$ is
significantly larger than the beaming angle of the jet,
$1/\Gamma$, at the considered time, the geometry is equivalent to
the isotropic case for observers inside the jet angle cone, and
the received fluence will not be depressed. For a weak burst like
GRB060218 and a typical stellar wind, this requires that the jet
opening angle is larger than $0.25$ radian, which may be
reasonable considering that the jet has a low energy. However, for
a jet with an opening angle comparable to  $1/\Gamma$, the
received fluence will be reduced by a factor of $\sim 8$.
Nontheless, in the event of GRB060218/SN2006aj, even considering
this effect the sub-GeV photon fluence could be detected by GLAST.

The anisotropic  IC scattering effect can reduce the emission
inside the $1/\Gamma$ cone of the jet,  but enhance the emission
at larger angles. This has potentially important implications for
detecting GeV emission from ``burstless" (untriggered)  hypernovae
during the early stage of their explosions, when the observer lies
off-axis respect to the GRB jet. The shock breakout soft X-ray
flash could still be  observed by wide-field X-ray cameras, such
as the proposed future {LOBSTER} mission, although they would not
trigger GRB detectors.

\acknowledgments XYW would like to thank  Z. G. Dai and Zhuo Li
for useful discussions. This work is supported by NASA NAG5-13286,
NSF AST 0307376, and  the National Natural Science Foundation of
China (for XYW).


\begin{thebibliography}{99}
\bibitem[]{530}
 Brunetti, G. 2000, Astroparticle Physics, 13, 107


\bibitem[]{534}
Calzavara, A. J. \&  Matzner, C. D., 2004, MNRAS, 351, 694
\bibitem[]{536}
Campana, S. et al. 2006, subm. Nature, astro-ph/0603279

\bibitem[]{539}
Colgate, S. A. 1974, ApJ, 187, 333

\bibitem[]{542}
Cusumano, G.  et al. 2006, GCN 4786.

\bibitem[]{545}
Dai, Z. G. \& Lu, T., 1998, MNRAS, 298, 87

\bibitem[]{548}
 Ensman, L. \& Burrows, A., 1992, ApJ, 393, 742

 \bibitem[]{551}
 Fan, Y. Z. \& Piran, T.,  2006, astro-ph/0601619

\bibitem[]{554}
Ghisellini, G. et al. 1991, MNRAS, 248, 14

\bibitem[]{557}
Klein, R. I. \& Chevalier, R. A. 1978, ApJ, 223, L109



\bibitem[]{562}
Mazzali, P. A. et al. 2006 subm. Nature,  astro-ph/0603567

\bibitem[]{565}
\Meszaros, P. \& Rees, M. J. 2001, ApJ, 556, L37



\bibitem[]{570}
 Modjaz, M. et al. 2006, subm. ApJ,  astro-ph/0603377



\bibitem[]{575}
 Pian, E. et al. 2006, subm. Nature, astro-ph/0603530

\bibitem[]{578}
Piran, T., 2005, Rev. Mod. Phys., 76, 1143

\bibitem[]{581}
Ramirez-Ruiz, E., Celotti, A. \& Rees, M. J., 2002, MNRAS, 337,
1349

\bibitem[]{585}
Ramirez-Ruiz, E. et al., 2005, ApJ, 625, L91

\bibitem[]{588}
Sazonov, S. Y., Lutovinov, A. A. \& Synyaev, R. A., 2004, Nature,
430, 646

\bibitem[]{592}
Soderberg, A. M. et al. 2004, Nature, 430, 648



\bibitem[]{597}
 Sollerman, J. et a. 2006, subm, A\&A, astro-ph/0603495
\bibitem[]{599}
Tan, J. C., Matzner, C. D. \& McKee, C., 2001, ApJ, 551, 946



\bibitem[]{604}
Watson, D. et al. 2004, ApJ, 605, L101



\bibitem[]{609}
Wang, X. Y., Li, Z. \& M\'esz\'aro, P., 2006, ApJ, 641, L89


\bibitem[]{613}
Waxman, E. \& Loeb, A., 2001, Phys. Rev. Lett., 87, 071101

\bibitem[]{616}
Waxman, E.  2004, ApJ, 602, 886

\bibitem[]{619}
Zhang, B., \& M\'esz\'aros, P. 2001, ApJ, 559, 110

\bibitem[]{622}
Zhang, B. \& \Meszaros, P., 2004, Int. Journ. of Mod. Phys. A, 19,
2385

\bibitem[]{626}
Zhang, W. Q., Woosley, S. E. \& MacFadyen, A. I., 2003, ApJ, 586,
356


\end{thebibliography}
\end{document}